\documentclass[12pt]{article}
\usepackage{graphicx} 

\usepackage{xcolor}
\usepackage{jheppub} 

\usepackage[T1]{fontenc} 

 \usepackage{bm}
 \usepackage{amsmath}
\usepackage{graphicx}
\usepackage{subfig}
\usepackage{multirow}
\usepackage{makecell}

\usepackage{amsmath}
\usepackage{braket}
\usepackage{soul}
\usepackage{slashed}
\usepackage{url}
\usepackage{hyperref}
\graphicspath{{./Figures/}}

\title{Limits on an Exotic Higgs Decay From a Recast 
ATLAS Four-Lepton Analysis}
\author{Junyi Cheng$^*$, Rabia Husain$^*$, Lingfeng Li$^\dagger$, Matthew J. Strassler$^*$}
\affiliation{* Department of Physics, Harvard University, Cambridge, MA, 02138, USA}
\affiliation{$\dagger$ Department of Physics, Brown University, Providence, RI, 02912, USA}
\date{November 2024}

\begin{document}

\abstract{The ATLAS collaboration, using 139 fb$^{-1}$ of 13 TeV collisions from the Large Hadron Collider, has placed limits on the decay of a $Z$ boson to three dark photons.  We reproduce the results of the ATLAS analysis, and  then recast it as a limit on a exotic Higgs decay mode, in which the Higgs boson decays via a pair of intermediate (pseudo)scalars $a$ to four dark photons $V$ (or some other spin-one meson).  Across the mass range for $m_a$ and $m_V$, we find limits on the exotic Higgs branching fraction BR$(H\to aa \to VVVV)$ in the range of $4\times 10^{-5}$ to $1 \times 10^{-4}$.}
\maketitle

\section{Introduction}

The discovery of the Higgs boson by the ATLAS and CMS
experiments in 2012 marked a watershed in high-energy physics.  Since then, active study of the production and decay modes of the Higgs boson have been underway at the Large Hadron Collider [LHC]~\cite{ATLAS:2022vkf,CMS:2022dwd}.  Many of the modes predicted in the Standard Model [SM] have been observed and measured.

However, the Higgs boson has a very narrow width, making it very sensitive to new interactions arising from physics beyond the Standard Model [BSM]. New BSM low-mass neutral particles that interact with the Higgs field can easily cause the Higgs boson to decay exotically with measurable rates.  In \cite{Curtin:2013fra}, a theoretical review was given of a wide variety of possible exotic Higgs decays with up to four prompt observable SM particles or partons.   Since then, searches have been carried out for many of these decay modes \cite{ATLAS:2021hbr,ATLAS:2018pvw,CMS:2018qvj,CMS:2018zvv,ATLAS:2020ahi,CMS:2020ffa,CMS:2021pcy,ATLAS:2021ldb,CMS:2022fyt,ATLAS:2023etl,ATLAS:2023ian,CMS:2024zfv,CMS:2024vjn,CMS:2024ibt,CMS:2024uru,ATLAS:2024vpj,ATLAS:2024wxf,ATLAS:2015hpr,ATLAS:2018coo}.

\begin{figure}[h]
  \centering
    \includegraphics[width=0.5\textwidth]{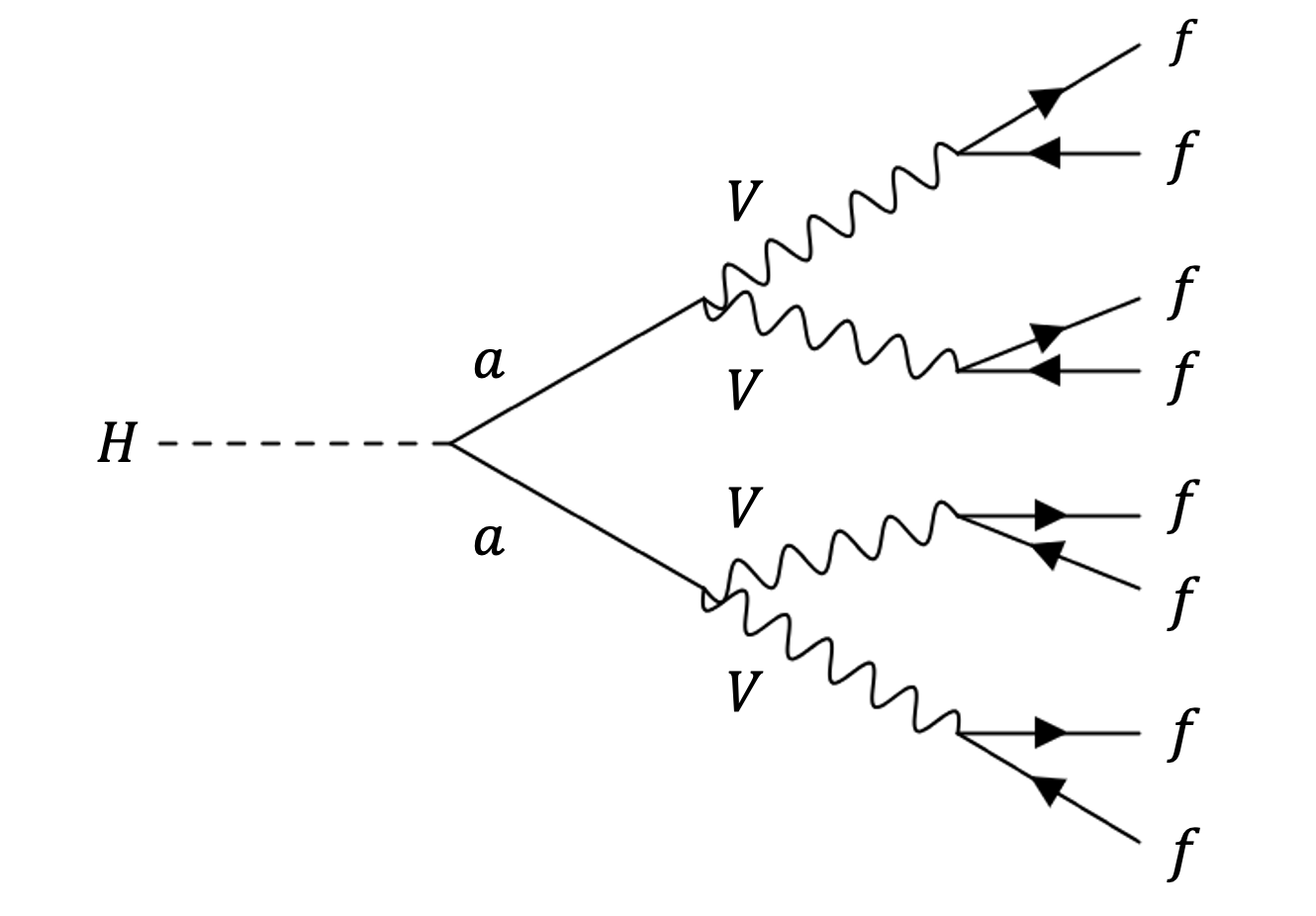}
    \caption{Feynman diagram for Higgs decay to 8 fermions}
    \label{fig:Higgs8f}
\end{figure}

In this paper, we consider instead an exotic Higgs decay mode with eight prompt SM fermions in the final state, shown in Fig.~\ref{fig:Higgs8f}. 
In this mode, the Higgs decays to four neutral spin-one particles $V$, via an intermediate scalar or pseudoscalar $a$, following which $V$ decays to a SM fermion-antifermion pair. We will refer to this mode as ``$H\to8f$.''  
When two or more of the $V$ bosons decay leptonically, the resulting signature consists of four or more prompt leptons in at least two same-flavor/opposite-charge (SFOC) pairs, with each SFOC pair having the same invariant mass $m_V$ and the four leptons having mass $m_{4\ell}<m_H-2m_V$. Such a signature has very low backgrounds, making it an excellent experimental target.

Despite its low backgrounds, this signature has not often been considered by the ATLAS or CMS experiments at the LHC. The exceptions usually have made additional kinematic requirements or limited themselves to low $V$ masses~\cite{ATLAS:2021yyr,CMS:2017tco,ATLAS:2015itk,CMS:2018jid, ATLAS:2015hpr, ATLAS:2018coo, ATLAS:2021ldb, CMS:2021pcy, CMS:2011xlr, CMS:2012qms, CMS:2015nay}.

\begin{figure}[h]
  \centering
    \includegraphics[width=0.5\textwidth]{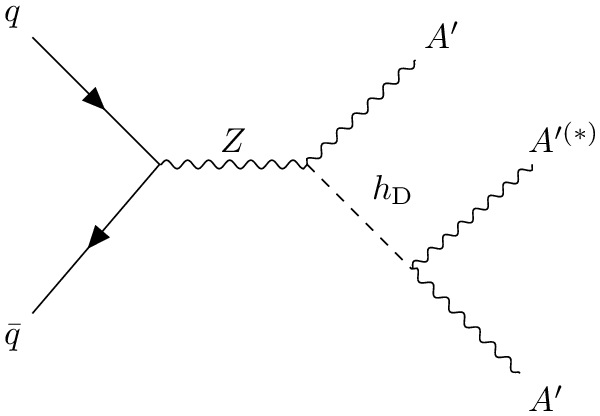}
    \caption{Feynman diagram for $Z$ decay to three $A^\prime$.}
    \label{fig:Z6f}
\end{figure}

However, ATLAS has recently published a search sensitive to $H\to8f$. The analysis \cite{ATLAS:2023jyp} targets a four-lepton final state that arises from a $Z$ boson decay to three dark photons $A'$, with a dark-sector Higgs $h_D$ as an intermediate final state, shown in Fig.~\ref{fig:Z6f}. We will refer to this mode as ``$Z\to6f$.''  The ATLAS analysis places limits on this exotic $Z$ decay mode as a function of the $h_D$ and $A'$ masses,  for $m_{A'}>5$ GeV.

In contrast to the powerful searches \cite{CMS:2021pcy,ATLAS:2021ldb,ATLAS:2015hpr,ATLAS:2018coo}   for the exotic decay mode $H\to VV\to (\ell^+\ell^-) (\ell^+\ell^-)$, which require the invariant mass of the four leptons $m_{4\ell}$ to be equal to $m_H$, the ATLAS search for $Z\to6f$ instead requires $m_{4\ell} < m_Z$.  With these reduced requirements on $m_{4\ell}$, this search can be recast as a limit on $H\to8f$ decays.  

A different approach to searching for $H\to8f$, based on counting leptons, was suggested by Izaguirre and Stolarski \cite{Izaguirre:2018atq}. A relevant search by ATLAS has been carried out \cite{ATLAS:2021yyr}, but it has not to our knowledge been recast for this decay mode. We will return to this issue in our concluding section.

Like the $Z\to6f$ decay mode, $H\to8f$ decays can easily arise in low-mass hidden sectors, often called ``hidden valleys'' \cite{Strassler:2006im} or ``dark sectors'' [HV/DS]. Such decays can appear perturbatively in theories with a hidden Higgsed $U(1)$, such as \cite{Schabinger:2005ei}, where $a$ can either be an elementary scalar or pseudoscalar, with a higher-dimension operator inducing the $a\to VV$ decay.  It can also arise non-perturbatively in a confining hidden sector; either or both $a$ and $V$ could be composite.  For instance, the role of $a$ could be played by a hidden $\pi$ meson, which could decay to elementary dark photons via an anomaly just as a QCD pion decays to SM photons.   However, the limits we place are largely model-agnostic, as discussed in Sec.~\ref{sec:H8f}.

 We will begin by reviewing the ATLAS search.  We next describe how we reproduce the ATLAS analysis, showing that we obtain results consistent with, and only slightly worse than, those obtained by ATLAS. We then recast this analysis to obtain limits on the exotic Higgs decay mentioned above, and  conclude with some discussion of how this measurement might be extended and improved.

In a companion paper in preparation, we will apply these methods to more general HV/DS models, which have confinement that differs from that of QCD.  In these confining hidden sectors, dark showering and dark hadronization often play a crucial role, and more careful consideration must be given to theoretical uncertainties.

\

\section{The ATLAS analysis}

We begin with a brief review of the ATLAS $Z\to 6f$ analysis~\cite{ATLAS:2023jyp}.  The following section will describe our efforts to reproduce its results.

\subsection{Triggers}

ATLAS 
uses a variety of  trigger streams in selecting events for this analysis.  These are the following, where the number(s) in parentheses indicate the minimum $p_T$(s) required:
\begin{itemize}
    \item single electron (26 GeV), 
    \item single muon (26 GeV),  
    \item dielectron symmetric (17 GeV),  
    \item dimuon symmetric (14 GeV),  
    \item dimuon asymmetric (22 GeV and 8 GeV),  
    \item electron (17 GeV) and muon (14 GeV),  
    \item two electron (12 GeV) and muon (10 GeV), and
    \item electron (12 GeV) and two muon (10 GeV). 
\end{itemize} 
All electrons must have $|\eta_e|<2.47$, while all muons must have $|\eta_\mu|<2.7$ \cite{ATLAS:2020gty,ATLAS:2019dpa}.

\subsection{Cut flow}
\label{ssec:cuts}

Events are selected using the following cut flow.  
 First, kinematic cuts are imposed on all observed leptons.  Electrons must have $p_T > 4.5$ GeV and $|\eta| < 2.47$.  Muons must have $p_T > 3$ GeV and $|\eta| < 2.7$.  All leptons must be prompt.\footnote{
  The longitudinal impact parameter $z_0$ relative to the primary vertex must satisfy $|z_0\, {\rm{sin}}(\theta)| < 0.5$, where $\theta$ is the angle of the lepton track relative to the beampipe.  The transverse impact parameter $d_0$, measured with uncertainty $\sigma_d$, must satisfy $d_0/\sigma_d < 3\ (5)$ for muons (electrons).}
 
 Leptons must also be isolated, with loose isolation requirements~\cite{ATLAS:2016lqx,ATLAS:2019qmc, ATLAS:2020auj}.  Electrons of transverse momentum $p_T^e$ are required to have $E_{\rm{cone}20} < 0.2\, p_T$ and $(p_T)_{\rm{varcone}20} < 0.15\, p_T$, where
 \begin{itemize}
     \item $E_{\rm{cone}20}$ is the energy of all particles within a cone of $\Delta R = 0.2$ surrounding the electron, and 
     \item $(p_T)_{\rm{varcone}20}$ denotes the scalar sum 
     of the transverse momentum (relative to the beam) of all {\it charged} particles (with $p^c_T >1$ GeV and $|\eta| < 2.5$) that lie within a cone of radius $\Delta R =$ min(10 GeV/$p_T^e$, 0.2) around the electron.
 \end{itemize}
 Muons are required to satisfy $(p_T)_{\rm{varcone}30} + 0.4\, E_{\rm{neflow}20} < 0.16\, p_T^\mu$, where
 \begin{itemize}
     \item $E_{\rm{neflow}20}$ is the transverse energy of all 
     {\it neutral} particle flow candidates within a cone of $\Delta R = 0.2$ surrounding the muon, and
     \item $(p_T)_{\rm{varcone}30}$ denotes the scalar sum of the transverse momentum (relative to the beam) of all {\it charged} particles (with $p^c_T >0.5$ GeV and $|\eta| < 2.5$) that lie within a cone of radius $\Delta R =$ min(10 GeV/$p_T^\mu$, 0.3) around the muon.

 \end{itemize}
If at least four leptons satisfy all of these requirements --- we will refer to this as the ``ID $\geq4$'' step -- the leptons are combined into same-flavor opposite-charge (SFOC) pairs in all allowed combinations.  If no combination has at least two SFOC pairs, the event is rejected. 

Any two  SFOC pairs forms a lepton quadruplet. 
 All possible SFOC quadruplets must have invariant mass $m_{4\ell}< m_Z -(5$ GeV), or the event is rejected.
 Within a quadruplet, the invariant masses of the SFOC pairs are labeled as $m_{12}$ and $m_{34}$, with $m_{12}>m_{34}$. If the SFOC assignment is ambiguous, as in events with 4 leptons of the same flavor, then the pairing is selected that has the smaller value of $m_{12}-m_{34}$.  Similarly, if there are multiple quadruplets, the one with the smallest $|m_{12}-m_{34}|$  is chosen.

 A cut is then placed on the angular separation between leptons in the quadruplet. Same-flavored leptons must have $\Delta R > 0.1$ and different-flavored leptons must have $\Delta R > 0.2$. 
 
 Since the signal includes two equal-mass lepton pairs, ATLAS requires that $m_{34}/m_{12} > 0.85$. Finally, to avoid background from hadronic resonances, all SFOC pairs in the quadruplet (including alternative pairings if the leptons are all of the same flavor) must satisfy 
 $$m_{ij}>5 \  {\rm{ GeV}} \ \ {\rm{and}} \ \ 
 (m_{\Upsilon(1s)} - 0.7\, {\rm{ GeV}}) < m_{ij} < (m_{\Upsilon(3s)} + 0.75\,{\rm{ GeV}}) \ ,$$  where $m_{\Upsilon(1s)} = 9.460$ GeV and $m_{\Upsilon(3s)} = 10.355$ GeV.

\subsection{Dark photon branching fractions}

Since dark photons appear in the ATLAS analysis and in some of the other signals used in this paper, we  briefly discuss how we treat dark photons.

\begin{figure}[h]
  \centering
    \includegraphics[width=0.6\textwidth]{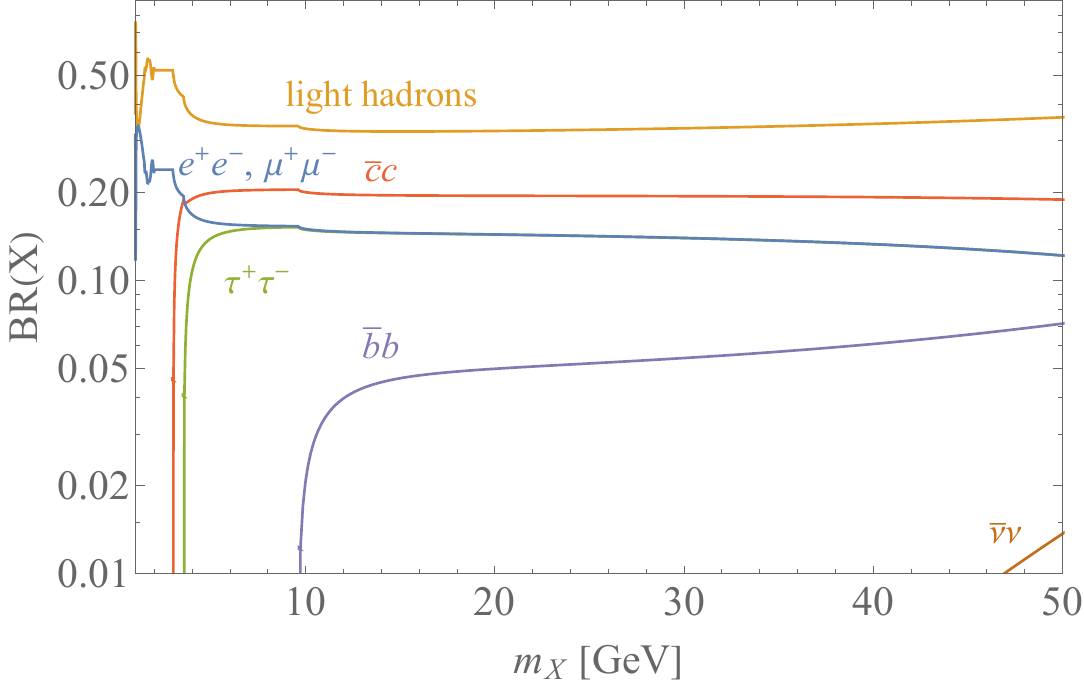}
    \caption{Branching ratios of the dark photon as a function of its mass. Here the ``light hadron curve'' stands for any hadronic final state with $u$, $d$, and $s$ quarks.}
    \label{fig:BR}
\end{figure}

Following the theory framework of \cite{Gopalakrishna:2008dv}, we introduce a $U(1)_D$  coupled to the SM through kinetic mixing between the new gauge boson $X_{\mu}$ and the hypercharge gauge boson $B_{\mu}$ (all quantities with a caret are the bare ones before diagonalization). The relevant terms in the Lagrangian are
\begin{equation}
    \mathcal{L}_X = -\frac{1}{4}\hat{X}_{\mu\nu}\hat{X}^{\mu\nu} + \frac{\chi}{2}\hat{X}_{\mu\nu}\hat{B}^{\mu\nu}-\frac{\hat{m}_X^2}{2} \hat{X}^\mu \hat{X}_\mu\ ,
\end{equation}
where $\hat{X}_{\mu\nu}$ $(B_{\mu\nu})$ is the field strength tensor of $\hat{X}_{\mu}$ ($B_\mu$), and $\xi$ is a small constant. We can diagonalize the kinetic terms and calculate the mass eigenstates of the massless photon, $Z$, and $X$. The branching ratio of the dark photon to various SM final states, calculated with DarkCast~\cite{Baruch:2022esd}, are shown in Fig.~\ref{fig:BR}.

\section{Reproducing the ATLAS analysis}

Our first task in recasting the ATLAS analysis~\cite{ATLAS:2023jyp} is to reproduce its results.  Aiding us in this effort is a Monte Carlo study that ATLAS provided as part of the supplemental material to the analysis, which we use to estimate the efficiencies and uncertainties that must be applied to truth-level simulated events to obtain limits on the signal cross-section~\cite{ATLAS:2023aux}.

\subsection{The ATLAS MC Study}
\label{ssec:Efficiency}

The ATLAS MC study includes the simulation of five points in parameter space, all with $m_{h_D}=50$ GeV.  We study the three cases $m_{A'} = 8, \ 15, \ 20$ GeV; the other two have off-shell dark photons ($m_{A'}>\frac12 m_{h_D}$) and lie outside our focus.

For the MC study, a simulation is done at 13 TeV center of mass, combining MadGraph 5\_aMC@NLO \cite{Alwall:2014hca} for the partonic process and PYTHIA 8.230 \cite{Sjostrand:2014zea} for showering and hadronization, using the A14 parton-shower tune \cite{TheATLAScollaboration:2014rfk}. The NNPDF3.0nlo parton distribution function is used \cite{NNPDF:2014otw}.  
Pile-up is not included.  The hard process $pp\rightarrow Z\rightarrow A'h_D$, $h_D\rightarrow A'A'$ was generated in MadGraph using the HAHM model \cite{Schabinger:2005ei, Curtin:2013fra,Curtin:2014cca} with $\epsilon=10^{-2}$, $\kappa=10^{-10}$, $m_{h_D}=50$ GeV, and 6 GeV $\leq m_{A'}\leq$ 24 GeV.    Branching ratios were computed automatically in MadGraph, and Madspin was used to allow the three $A'$s to decay to SM fermion-antifermion pairs.  

The ATLAS MC study takes a shortcut that ensures that all events have at least four signal leptons (here lepton means electron or muon.) 
This is done by letting one of the $A'$ decay into any kinetically allowed fermion pair, including quarks and neutrinos, while restricting the decay of the other two $A'$ to $ee$ or $\mu\mu$. The contribution of decays to $\tau\tau$ was ignored, as the analysis cuts remove most such events.\footnote{A matching procedure is adopted if $m_{A'}\geq\frac{1}{2}m_{h_D}$ or $m_{A'}+m_{h_D}\geq m_Z$.  We ignore this issue here, since we only focus on the part of parameter space where $h_D$ and all $A'$s are on-shell.}

This shortcut only approximately reproduces the count of 6 lepton events versus 4, as the combinatorics of leptonic decays is not strictly correct. This results in a small error in the ATLAS MC study simulation relative to a full simulation of the model.  In using this ATLAS MC study to estimate efficiencies, we use the same, slightly erroneous, method. However, in reproducing the full analysis and in imposing limits on other models, we do complete simulations, avoiding this approximation. 

One additional detail is that the ATLAS Monte Carlo study imposes the trigger {\it after} the cut flow.  This unrealistic ordering means that, after having matched the ATLAS MC study, we have to estimate the uncertainties in the efficiencies arising from the reversed order. Fortunately these appear small; see Section~\ref{sec:trigger first}.

The cut flow of the MC study mostly follows the flow in the analysis as described above in Section~\ref{ssec:cuts}, although it begins by imposing a ``Monte Carlo (MC)'' cut that removes all simulated events except those that have at least four leptons with $p_T > 2$ GeV and $\eta < 3$.  Cuts are then imposed in the order described above in Section~\ref{ssec:cuts}.

\subsubsection{Reproducing the ATLAS MC Study}

We have repeated this MC study, following the ATLAS procedure carefully, with MadGraph to simulate the partonic events and PYTHIA for showering and hadronization. All settings were the same as for ATLAS's study, except that we use PYTHIA 8.308.  
At truth-level, the efficiencies of the cuts, relative to the cuts that precede them, are shown in Table~\ref{table:truth}. Following ATLAS, the ID $\geq 4$ cut at the second step, in which the number of identified leptons must be $\geq 4$, includes the requirement that the leptons be isolated as described in Sec.~\ref{ssec:cuts}. The $\Delta$R cut is included in the SFOC step of the cut flow.\footnote{When matching these efficiencies, note than an event with no quadruplets {\it passes} the $m_{4l}$ cut; it is then rejected by the SFOC$\geq 2$ requirement.}

 While ATLAS has not published a table for these generator-level efficiencies, discussions about the MC study with its authors  \cite{privatecommunication} confirms that our results are always close to those found by ATLAS.  Specifically, the 21 entries in Table~\ref{table:truth} differ usually by less than 4\% (relative), and at most by 8\% (relative), from the ATLAS simulations.  The overall signal efficiency, the most crucial number, differs by at most 8\% (relative) at the three mass points.

\begin{table}[!ht]
    \centering
    \caption{The truth-level efficiencies of our study at $m_{h_D} = 50$ GeV. The ID $\geq 4$ step requires four leptons that pass kinematic and isolation cuts; see Sec.~\ref{ssec:cuts} for details.}
    \label{table:truth}
    {\begin{tabular}{|l||c|c|c||}
    \hline
        $m_{A'}$  &   8 GeV & 15 GeV &  20 GeV \\
        \hline
        MC filter efficiency &  57.7$\%$ &  61.9$\%$ &  64.4$\%$ \\ \hline
        ID $\geq$ 4 & 54.2$\%$ &  53.5$\%$ & 56.7$\%$ \\ \hline
        $m_{4\ell} < m_Z - 5$ GeV &  97.9$\%$ &  98.9$\%$ &  99.4$\%$ \\ \hline
        \# SFOC lepton pairs $\geq$ 2 &  84.0$\%$ &  85.3$\%$ &  87.8$\%$ \\ \hline
        $m_{\ell_3\ell_4}/m_{\ell_1\ell_2} > 0.85$ & 95.1$\%$ &  95.3$\%$  & 96.6$\%$ \\ \hline
        No $m_{\ell^+\ell^-}<$ 5 GeV or near $m_\Upsilon$ &  95.3$\%$ &  92.3$\%$ &  90.9$\%$ \\ \hline \hline
        Overall signal efficiency &  23.3$\%$ &  24.6$\%$ &  28.0$\%$ \\ \hline
    \end{tabular}}
\end{table}

After detector effects are accounted for, ATLAS's cut flow efficiencies are shown in Table~\ref{table:reco}, along with the post-cut-flow trigger efficiency.  In our efforts to reproduce these results, we have chosen not to use a detailed detector simulation. Instead, we use a simplified approach, applying constant recalibration factors for lepton reconstruction and for triggering, and assigning substantial systematic uncertainties to those factors.

Specifically, we find that by imposing a constant recalibration factor $r_{\rm lep}=0.78$ per lepton after the ID $\geq 4$ cut, we reproduce the ATLAS MC study's pre-trigger efficiencies surprisingly well, as shown in Table~\ref{table:reco}. 
The success of a constant $r_{\rm lep}$ in reproducing the Monte Carlo study  suggests that our main failure to emulate the detector occurs in the ID $\geq 4$ cut. 
Once this is accounted for, there do not appear to be other major discrepancies. 

We also implement the triggers naively, treating them as though they were sharp truth-level cuts on the leptons' $p_T$ and $\eta$.  We must then correct our post-cut-flow trigger efficiency to match that of the MC study.  Another constant  recalibration  factor $r_{\rm trig}=0.81$ brings our trigger efficiencies in line with those of the study, as again shown in Table~\ref{table:reco}.

Of course, there is nothing justifiable about this crude methodology other than its success. 
In reality, efficiencies differ for electrons and muons, and depend on $p_T$ and $\eta$.  Moreover, we are combining identification efficiencies with isolation efficiencies (and indeed, since the ATLAS MC study itself combines them into a single cut, we have insufficient information to do better.)  Similarly, different triggers ought to have different recalibration factors.  Nevertheless, we believe this is the best we can do with the information we have. The leptons in this analysis are often soft, making lepton reconstruction, fake leptons, and lepton isolation impossible for us to model. In addition, published information from ATLAS on lepton efficiencies does not sufficiently cover soft and fake leptons. And finally, we have only information about the total trigger efficiency, not the efficiency for each trigger pathway or its turn-on curve.

As we attempt to reproduce the full ATLAS analysis, we will apply the two recalibration factors $r_{\rm lep}$ and $r_{\rm trig}$, along with our naive model of the triggers,  to our truth-level simulations.
 Our approach to quantifying the large uncertainties from this procedure is discussed in Section~\ref{ssec:statistics} below.

\begin{table}[!ht]
    \centering
    \caption{The reconstruction-level efficiencies of the ATLAS MC study at $m_{h_D} = 50$ GeV, compared with our own  after correcting our results with recalibration factors $r_{\rm lep}=0.78$ per lepton and $r_{\rm trig} = 0.81$.  The ID $\geq 4$ step requires four leptons that pass kinematic and isolation cuts; see Sec.~\ref{ssec:cuts}.}
    \label{table:reco}
    \resizebox{\columnwidth}{!}{\begin{tabular}{|l||c|c||c|c||c|c||}
    \hline
        $m_{A'}$  &  8 GeV & 8 GeV & 15 GeV & 15 GeV & 20 GeV & 20 GeV \\
        $ $ & ATLAS  & Our  & ATLAS  & Our  & ATLAS  & Our \\ 
        & Result  & Result  & Result  & Result  &Result  & Result   \\ \hline
        MC filter efficiency & 58.0$\%$ & 57.7$\%$ & 62.2$\%$ & 61.9$\%$ & 64.5$\%$ & 64.3$\%$ \\ \hline
        ID $\geq$ 4 & 27.2$\%$ & 27.1$\%$ & 26.9$\%$ & 27.1$\%$ & 28.4$\%$ & 29.1$\%$ \\ \hline
        $m_{4\ell} < m_Z - 5$ GeV & 96.9$\%$ & 98.4$\%$ & 98.0$\%$ & 99.2$\%$ & 98.8$\%$ & 99.6$\%$ \\ \hline
        \# SFOC lepton pairs $\geq$ 2 & 73.1$\%$ & 72.6$\%$ & 74.4$\%$ & 73.8$\%$ & 77.6$\%$ & 76.1$\%$ \\ \hline
        $m_{\ell_3\ell_4}/m_{\ell_1\ell_2} > 0.85$ & 86.2$\%$ & 89.5$\%$ & 86.7$\%$ & 90.6$\%$ & 87.4$\%$ & 92.4$\%$ \\ \hline
        No $m_{\ell^+\ell^-}<$ 5 GeV or near $m_\Upsilon$ & 92.0$\%$ & 95.2$\%$ & 91.7$\%$ & 92.4$\%$ & 90.1$\%$ & 91.0$\%$ \\ \hline \hline
        Trigger & 70.0$\%$ & 66.8$\%$ & 62.2$\%$ & 63.6$\%$ & 59.2$\%$ & 59.4$\%$ \\ \hline \hline
        Overall signal efficiency & 6.2$\%$ & 6.3$\%$ & 6.2$\%$ & 6.4$\%$ & 6.5$\%$ & 7.1$\%$ \\ \hline
    \end{tabular}}

\end{table}

\subsubsection{Trigger/Cut Flow Ordering \label{sec:trigger first}}

As noted above, the ATLAS MC study  applies the cut flow before the trigger. In recasting the full analysis, where the order is of course reversed, it is not at all obvious that the our method for correcting the trigger and cut flow, which we obtain from the ATLAS MC study, will still apply.      

To investigate this, we repeated the MC study putting the trigger and cut flow in a realistic order, while maintaining the same recalibration factors as before ($r_{\rm trig} =81\%$ and $r_{\rm lep}=78\%$.)  The resulting trigger efficiency is of course much lower, while the cut flow efficiencies are higher.  But the overall efficiencies, which for $m_{h_D}=50$ GeV vary from 0.060 at $m_V=6$ GeV to 0.076 at $m_V = 24$ GeV, are unaltered, differing by less than 1\% (relative). This indicates the trigger and cut flow largely commute, and so the trigger recalibration factor we obtained earlier can be be applied to an analysis of the real data, with no substantial uncertainty from this source.

\subsubsection{Pileup}

The ATLAS analysis provides us with no information about pileup through its Monte Carlo study.  Meanwhile, we cannot directly  simulate either pileup or pileup subtraction accurately. We have therefore not attempted to model it.

Fortunately, our study is mostly insensitive to pileup except in two aspects: pileup must be subtracted effectively from lepton isolation cones, and it must not contaminate electron $p_T$ measurements.  We will make the assumption that pileup subtraction in the ATLAS event sample is sufficiently effective that the ATLAS Monte Carlo study of pileup-free simulated data gives us a good measure of the efficiencies in real, pileup-subtracted data. More specifically, we assume that even though we cannot quantify the uncertainties from this issue, they are small compared to the quantifiable uncertainties that we have already described from other sources.

\subsection{Our Method for Obtaining Limits}
\label{ssec:statistics}
In this section, we will elaborate the procedures to obtain our limits on the signal strength $\mu_{\rm sig}\equiv {\cal L}\ \sigma(pp \to Z \to A^\prime h_D \to 4\ell +X)$, where ${\cal L}$ is the integrated luminosity of 139 fb$^{-1}$.  We produce truth-level samples as for the MC study (except that we simulate the full model rather than just four-lepton events.)   We apply the trigger pathways as cuts, along with the recalibration factor $r_{{\rm trig}}$ on events with four or more leptons, to set our initial event sample.  We then  apply the cuts of the ATLAS analysis to those that remain, keeping only a fraction $r_{\rm lep}$ of the leptons that pass the ID $\geq 4$ cut.

We then
 follow the approach in the ATLAS search~\cite{ATLAS:2023jyp}, dividing the signal region into multiple bins of $\bar{m}_{\ell\ell}\equiv (m_{\ell_1\ell_2}+m_{\ell_3\ell_4})/2$ with 1~GeV width. Each relevant $\bar{m}_{\ell\ell}$ bin is treated as an independent channel. In particular, for the $i$-th signal bin of a certain $\{m_{h_D},m_{A^\prime}\}$ benchmark, its average signal yield $S_i(m_{h_D},m_{A^\prime})$ is determined by
\begin{equation}
    S_i(m_{h_D},m_{A^\prime}) = \mu_{\rm sig} \times \epsilon_{\rm sig}(m_{h_D},m_{A^\prime}) \times \kappa_i(m_{h_D},m_{A^\prime})~,
\end{equation}
with
\begin{equation}
    \sum_{i=1}^{N_d} \kappa_i(m_{h_D},m_{A^\prime}) = 1~.
\end{equation}
Here $\epsilon_{\rm sig}(m_{h_D},m_{A^\prime})<1$ is the overall signal efficiency, $\kappa_i(m_{h_D},m_{A^\prime})$ is the normalized shape factor of the dilepton resonance peak, and $N_d$ is the total number of relevant $\bar{m}_{\ell\ell}$ bins. For brevity, we will omit the dependence of $S_i$, $\epsilon_{\rm sig}$ and $\kappa_i$ on the model parameters $m_{h_D}, m_{A^\prime}$ in all following equations, and (unless otherwise specified) expressions should be understood to implicitly account for these parameters. Meanwhile, the corresponding background yield $B_i$ for each channel is directly provided in Ref.~\cite{ATLAS:2023jyp}. 
Following the $CL_S$ method~\cite{Junk:1999kv,Read:2002hq,Cowan:2010js} widely accepted by the LHC community, the observed and expected limits on $\mu_{\rm sig}$ are then obtained once $\epsilon_{\rm sig}$ and $\{\kappa_i\}$ are known. 

As shown in Sec.~\ref{ssec:Efficiency}, the $\epsilon_{\rm sig}$ of the three benchmark points in Table~\ref{table:reco} match with the results of the ATLAS's MC study  once the necessary recalibration factors $r_{\rm lep},~r_{\rm trig}$ are applied.  With the same set of recalibration factors, we calculate $\epsilon_{\rm sig}$ for all other mass values. 

The signal shapes $\left\{\kappa\right\}$ are provided at multiple values of $m_{h_D},m_{A'}$ in the supplementary material of~\cite{ATLAS:2023jyp}.  We interpolate between these shapes to obtain the signal shape at other parameter points. 
Thanks to the excellent detector resolution on lepton momenta, the reconstructed $\bar{m}_{\ell\ell}$ peaks are narrow, allowing us to take the four leading $\bar{m}_{\ell\ell}$ bins to calculate $\kappa_i$ by normalizing their sum to unity. 
However, the signal region is always dominated by at most two $\bar{m}_{\ell \ell}$ bins close to $m_{A^\prime}$. We keep only the two leading $\bar{m}_{\ell \ell}$ bins in the following analysis to reduce computational cost.

Our methods rest on poorly known and crudely defined recalibration factors, so the systematic uncertainties on our estimates of $\epsilon_{\rm sig}$ are large.  We also share ATLAS's systematic uncertainty on the backgrounds $B_i$. It is clear that systematic uncertainties were important for ATLAS,\footnote{Were we to account only for statistical uncertainties, our limits would be stronger than those obtained by ATLAS.} and we must account for them as well.
 
 We introduce the systematic effects by promoting the overall signal efficiency $\epsilon_{\rm sig}$ and the expected background yield $B_i$ in the $i$-th bin to nuisance parameters instead of constants. The true distributions are unknown, but here we will approximate them  by log-normal distributions.
 We also assume that, for each model benchmark $\{m_{h_D},m_{A^\prime}\}$, the uncertainties in the two $\bar{m}_{\ell\ell}$ bins that we retain are strongly correlated, so that they can be approximately described by the same parameter. 
 
 We then introduce parameters $\delta_{\epsilon}$ and $\delta_B$ for each $\{m_{h_D},m_{A^\prime}\}$ benchmark, defined as follows:
\begin{equation}
    \epsilon_{\rm sig} =  \bar{\epsilon}_{\rm sig} \times (1+\delta_{\epsilon})~,~ B_i =  \bar{B}_i \times (1+\delta_{B})~,
\end{equation}
and
\begin{equation}\label{eq:lognormal}
    \log (1+\delta_{\epsilon}) \sim \mathcal{N}(0,\sigma_\epsilon^2)~,~\log (1+\delta_{B}) \sim \mathcal{N}(0,\sigma_{B}^2)~,
\end{equation}
where $\bar{\epsilon}_{\rm sig}$ and $\bar{B}_i$ are the central values obtained from the previous procedure.
Even when $\sigma_{\epsilon}$ or $\sigma_{B}$ is $\mathcal{O}(1)$, the distribution in Eq.~(\ref{eq:lognormal}) ensures the expected event yields are positive.

The SM background distributions ${\bar B}_i$ and systematic uncertainties $\sigma_{B,i}$ are provided in the ATLAS analysis~\cite{ATLAS:2023jyp}.  The quantity $\sigma_\epsilon$ for each mass benchmark  can be decomposed into several independent terms:
\begin{equation}
\label{eq:sigmaepsilon}
    \sigma_\epsilon^2 \simeq \sigma_{\rm trig}^2 + (4\sigma_{\rm lep})^2+ \sigma_{\rm theo}^2~, 
\end{equation}
where $\sigma_{\rm trig}$, $\sigma_{\rm lep}$ are the uncertainties of the corresponding recalibration factors  $r_{\rm trig}$, $r_{\rm lep}$ obtained in Sec.~\ref{ssec:Efficiency}, and $\sigma_{\rm theo}$ is a theoretical uncertainty.\footnote{Note that the lepton identification uncertainties for the four leptons are strongly correlated and thus appear in Eq.~(\ref{eq:sigmaepsilon}) collectively.}  We now explain how we estimate these uncertainties.

The two recalibration factors characterize the discrepancy between our simplified simulation, which does not include detector efficiencies and trigger thresholds, and a more sophisticated simulation.  But our success in matching the MC study suggests a method for estimating the uncertainties in these two factors.  Neither would be expected to be close to one, and of course neither can exceed one.  This motivates us to assign the 1$\sigma$ fluctuation of $\epsilon_{\rm sig}$ to be $\frac{1}{2}(1-r_{\rm sig})=0.11$, which puts the unreasonable circumstance of $r_{\rm sig}>1$ a full 2$\sigma$ away from the central value.\footnote{
This uncertainty is comparable to the difference between our $r_{\rm lep}$ value and ATLAS's average lepton efficiency in this $p_T$ range, as one can infer from ATLAS technical reports~\cite{ATLAS:2016lqx,ATLAS:2019qmc}. } The corresponding $\sigma_{\rm lep}$ is therefore $(1-r_{\rm lep})/2 r_{\rm lep}=0.14$. Similarly, we take $\sigma_{\rm trigger}$ to be $(1-r_{\rm trig})/2 r_{\rm trig}=0.12$.

The term $\sigma_{\rm theo}$ denotes theoretical uncertainties induced by perturbative calculations, hadronization, and parton distribution functions. Excluding any detector effects, the simulations described in Sec.~\ref{ssec:Efficiency} are almost identical to the ones in the ATLAS $Z\to 6f$ study.   We therefore set this term to be $\sigma_{{\rm theo}}=0.14$, the same as in the ATLAS analysis. 

Taking this into account, the overall $\sigma_{\epsilon}$ is 0.59, conservative but not unreasonable for a theoretical study.  Its size makes clear why we have chosen log-normal distributions rather than Gaussian distributions for our analysis.   

There are other contributors to $\sigma_{\epsilon}$ not included above, such as effects from pile-up and from detector energy resolution. However, according to~\cite{ATLAS:2023jyp}, the size of such ``experimental uncertainties'' (in which pile-up is included) on signal efficiency is at most $7\%$, too small to impact the above value of $\sigma_\epsilon$. 
We thus do not include these terms for simplicity.

\begin{figure}[h!]
\centering
\includegraphics[width=.45\textwidth]{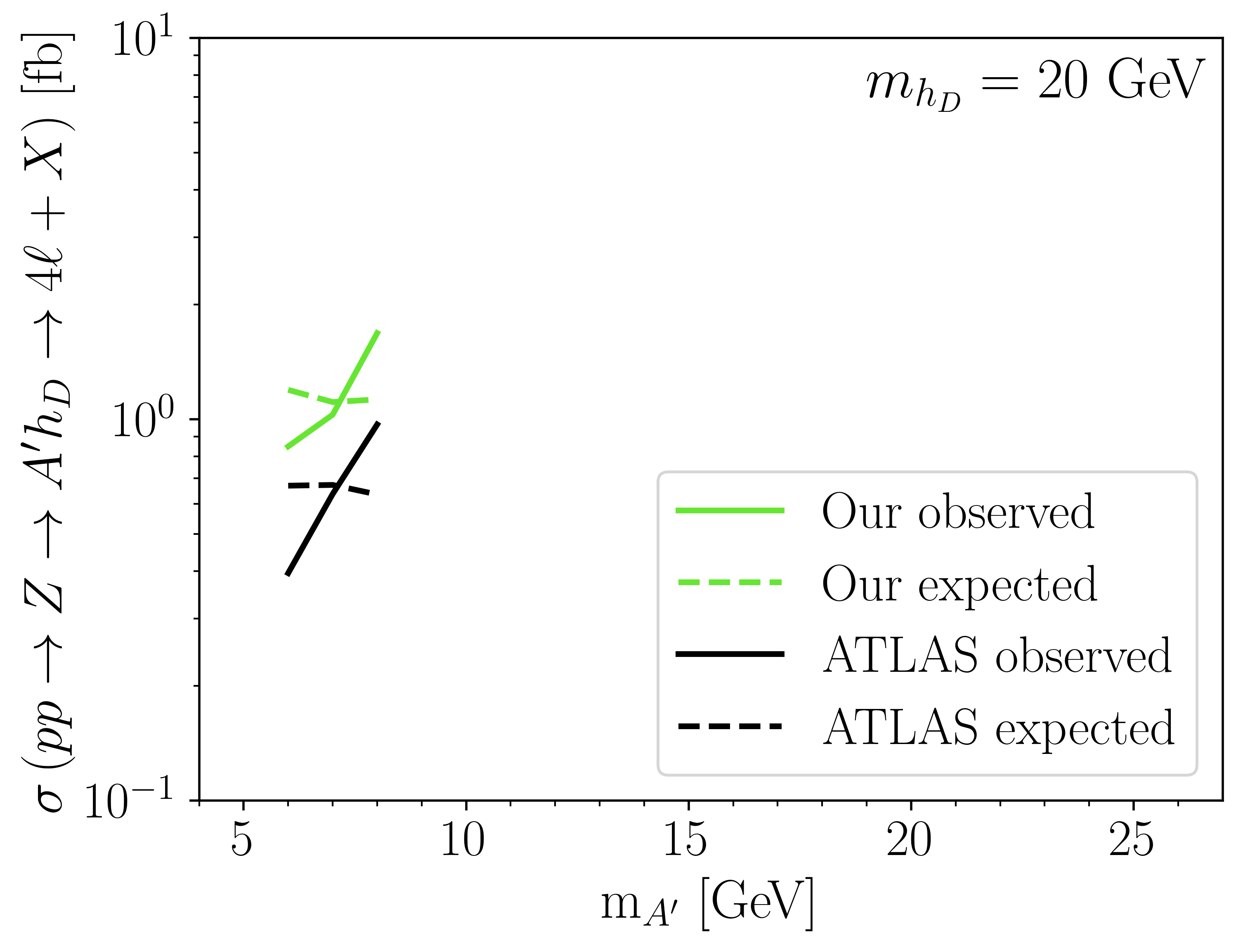}
\includegraphics[width=.45\textwidth]{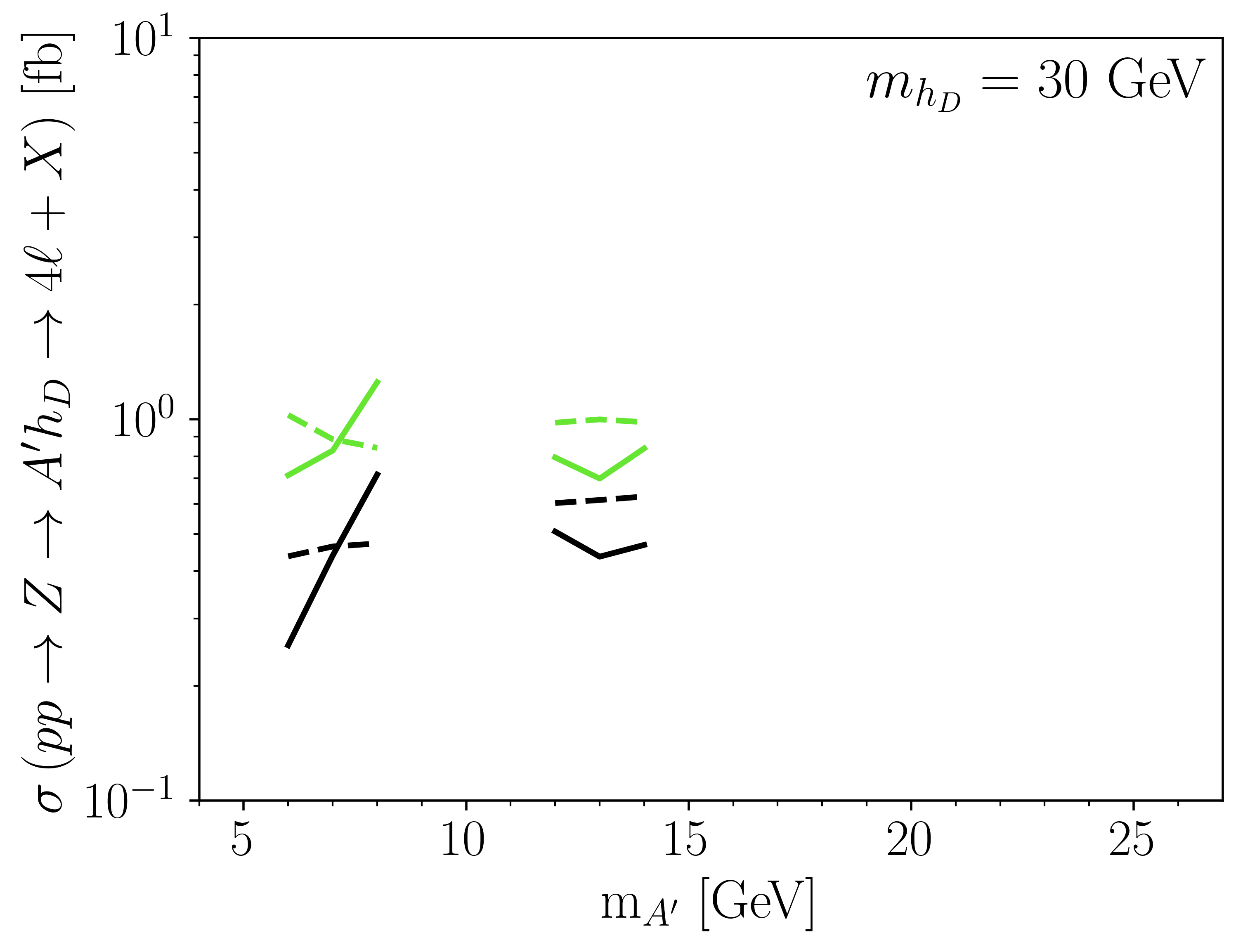}
\includegraphics[width=.45\textwidth]{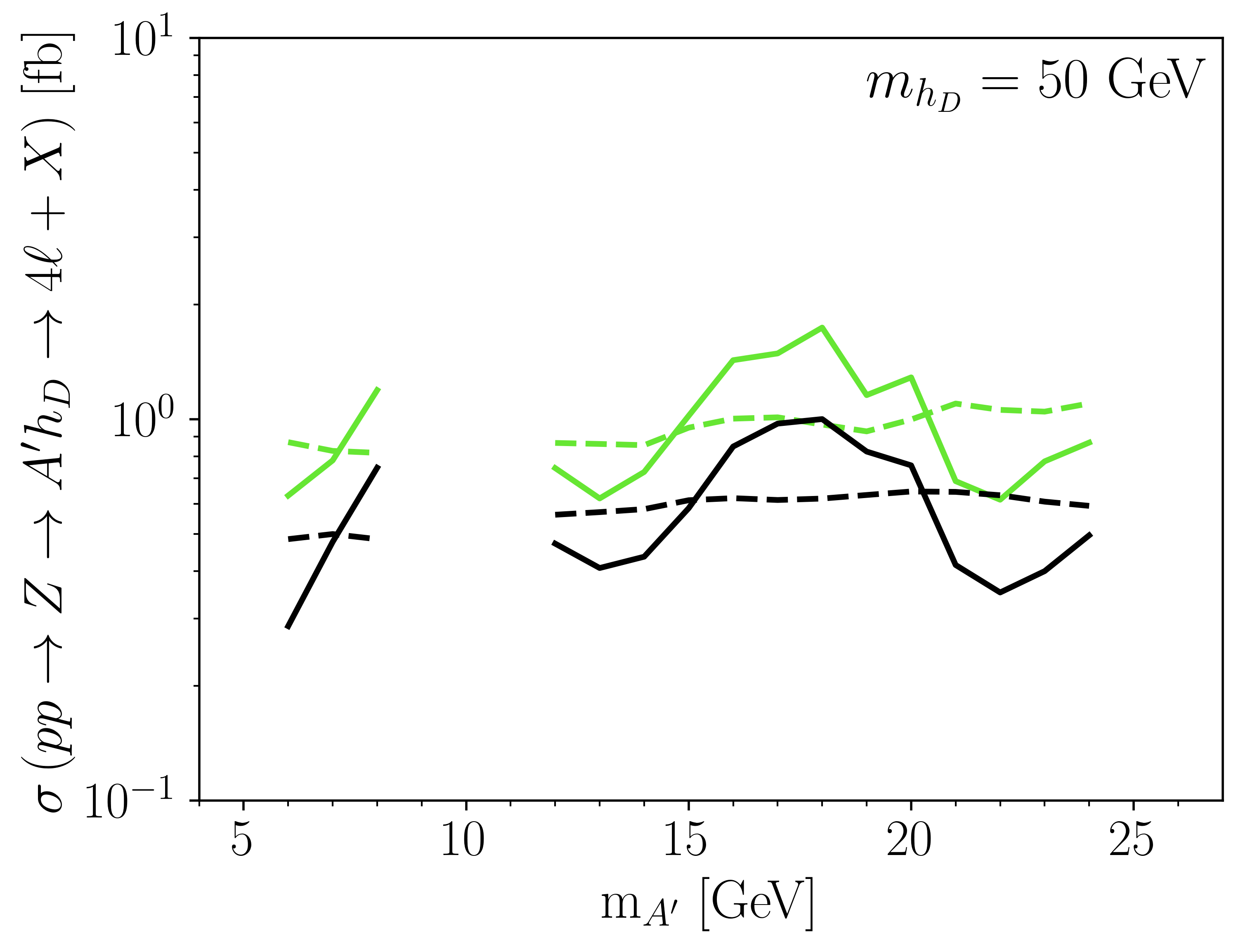}
\caption{Limits at 95\% confidence level on the cross section times branching fraction of $Z\to h_DA'\to A'A'A'$, where at least two $A'$ decay to $\ell^+\ell^-$.   ATLAS observed (expected) limits are shown as solid (dashed) black lines, while our observed (expected) limits are shown as solid (dashed) green lines.}
\label{fig:Z4l1}
\end{figure}

With systematic effects modeled in this way, we obtain 95\% C.L. limits on $\sigma(pp \to Z \to A^\prime h_D \to 4\ell +X)$  for various $\{m_{h_D},m_{A^\prime}\}$ benchmarks where the $h_D \to A^\prime A^\prime$ decay is on-shell. (Note that this limit is on four-lepton final states, not on three-$A'$ final states.)  For $m_{h_D} = 20, 30, 50$ GeV, we plot our expected and observed limits in Fig.~\ref{fig:Z4l1}, together with the those of ATLAS.\footnote{We do not place limits for $m_{A'} = 8,12$.  These are adjacent to the $\Upsilon$ mass cut, making it difficult for us to determine the signal shape $\left\{\kappa_i\right\}$.}   

Our results match the shape of ATLAS's limits, reproducing the $m_{h_D}$ and $m_{A'}$ dependence in both expected and observed limits. Our bounds are weaker by a factor that varies mainly between 1.5 and 1.9, except at the smallest mass $m_{A'}=6$ GeV where it tends higher.  This is not surprising given the crude nature of our event simulation, as captured in the large systematic errors on our recalibration factors.

A minor issue is that our recast observed limits tend to fluctuate more than the ATLAS ones, which is most obvious in the $\{m_{h_D},m_{A^\prime}\}=\{50,19\}$~GeV benchmark point. Since the expected limits do not experience similar fluctuations, the cause probably lies in the likelihood distribution of observed events. The statistical interpretation we adopt takes the background event count per $\bar{m}_{\ell\ell}$ bin and only accounts for the resolution via the signal shape. The potential correlation of background event counts across nearby $\bar{m}_{\ell\ell}$ bins, due to detector resolution, cannot be extracted from the public data, and this  may leave us more sensitive to bin-to-bin fluctuations than is the case for ATLAS. 

This largely successful reproduction of the results of the ATLAS analysis now gives us confidence to proceed to recasting the analysis for other BSM signatures, in particular for the $H\to 8f$ decay mode.

\section{Limits on the $H\to8f$ Decay Mode} 
\label{sec:H8f}

In the exotic $H\to8f$ decay mode of Fig.~\ref{fig:Higgs8f},  the Higgs decays to two scalars or pseudoscalars $a$, which in turn decay to two spin-one vectors $V$, each of which decays to lepton pairs a certain fraction of the time. 
As noted above, this signature can arise in many hidden valley/dark sector models. The simplest is the hidden abelian Higgs model \cite{Schabinger:2005ei, Curtin:2014cca} and multi-Higgs variants thereof, but it can also arise in strongly interacting HV/DS models, where $a$ or $V$ might be composite.  But this Higgs decay signature is largely agnostic to the details of the model --- for instance, it is hardly affected by the parity of $a$ or by whether $a$ or $V$ is elementary --- and so we will not discuss model specifics further. 

The only way in which model-dependence enters in a significant way into our limits is in the branching fraction of $V\to\ell\ell$.  We will assume $V$ has the branching fractions of a dark photon, shown in Fig.~\ref{fig:BR}.  (We also assume BR$(a\to VV)=100\%$.) For cases where BR($V\to\ell\ell$) is roughly similar to that of a dark photon, it is fairly straightforward to reinterpret our results, though the presence of six- and eight-lepton events means that one cannot simply rescale our limits. Nevertheless, as long as four-lepton events dominate, reasonable limits can be obtained by rescaling our limits by the square of the ratio of leptonic branching fractions for $V$ versus that of a dark photon.

We therefore treat   ${\rm BR}(H\to aa\to VVVV)$ as an unknown on which we place a limit. 
We simulate $H\rightarrow aa\rightarrow VVVV\rightarrow8f$ in PYTHIA 8.311  \cite{Sjostrand:2014zea}, including all relevant Higgs production processes.   We reweight their cross-sections to match the cross-sections recommended by the LHC Higgs working group \cite{Karlberg:2024zxx}. 
The branching ratios of $V$ decaying to various fermion-antifermion pairs are computed as described in Section \ref{ssec:Efficiency}.

We then recast the analysis~\cite{ATLAS:2023jyp} using the same methodology employed in the previous section.   From simulation, we find
the overall lepton $p_T$ distributions are similar to those in the $Z \to 6f$ case; note that $m_Z/6\approx m_H/8$, so this is roughly as expected.
We also note that the average lepton $|\eta|$ from $H\to 8f$ events is slightly smaller than that from $Z\to 6f$, due to different production mechanisms, which may lead to a slightly higher lepton efficiency at the detector level. For simplicity, we take the recalibration factors and their uncertainties  be the same as for $Z\to 6f$; see Sec.~\ref{ssec:statistics}. 

However, the theoretical uncertainty $\sigma_{\rm theo}$ term needs to be reevaluated, since the simulation of Higgs boson production using PYTHIA 8 is known to be inaccurate. Most importantly for our purposes, it is not suitable for computing the high-$p_T$ tail from $gg\to H$. 
The Higgs $p_T$ distribution in $gg\to H$ has been calculated at NNLO in \cite{Chen:2018pzu}. We find the Higgs $p_T$ distribution from PYTHIA 3.811 to be surprisingly similar to the results of \cite{Chen:2018pzu}, with a minor excess in the high $p_T$ region.  

Any mismodeling of the Higgs $p_T$ distribution may shift the central value of the total signal efficiency $\epsilon_{\rm sig}$ and affect our theoretical  uncertainty $\sigma_{\rm theo}$.  To study this, we first bin our truth-level Higgs events in bins of 25 GeV and calculate the average signal efficiency $\epsilon_{\rm sig}(\bar p_T)$ in each $p_T$ bin centered on $p_T=\bar p_T$. We can then estimate the total efficiency of any other Higgs $p_T$ distribution by binning it similarly and weighting it by $\epsilon_{\rm sig}(\bar p_T)$. Comparing the $p_T$ distribution from our PYTHIA 8 simulation and the one from~\cite{Chen:2018pzu}, we find the central value shift in the total efficiency $\epsilon_{\rm sig}$ is extremely small, much smaller than the uncertainties from our recalibration factors, and so we neglect it.  However, the uncertainty from this procedure is not small. Acounting both for the statistical uncertainty on our PYTHIA simulation, which dominates, and the intrinsic theoretical uncertainty in~\cite{Chen:2018pzu}, we find a theoretical uncertainty on $\epsilon_{\rm sig}$ from the Higgs kinematic distribution of about 15\%. We include this in our limit calculation.

There is also a theoretical uncertainty on the SM Higgs production rate~\cite{LHCHiggsCrossSectionWorkingGroup:2016ypw} of approximately $\sim 5\%$, arising mainly from the perturbative uncertainty on $gg\to H$, the value of $\alpha_s$, and uncertainties on the PDFs. Combining this with the uncertainty from the Higgs $p_T$ distribution leads to a final $\sigma_{\rm theo}$ of 0.16, which we combine with other uncertainties as in Eq.~(\ref{eq:sigmaepsilon}).

\begin{figure}[h!]
\centering
\includegraphics[width=.45\textwidth]{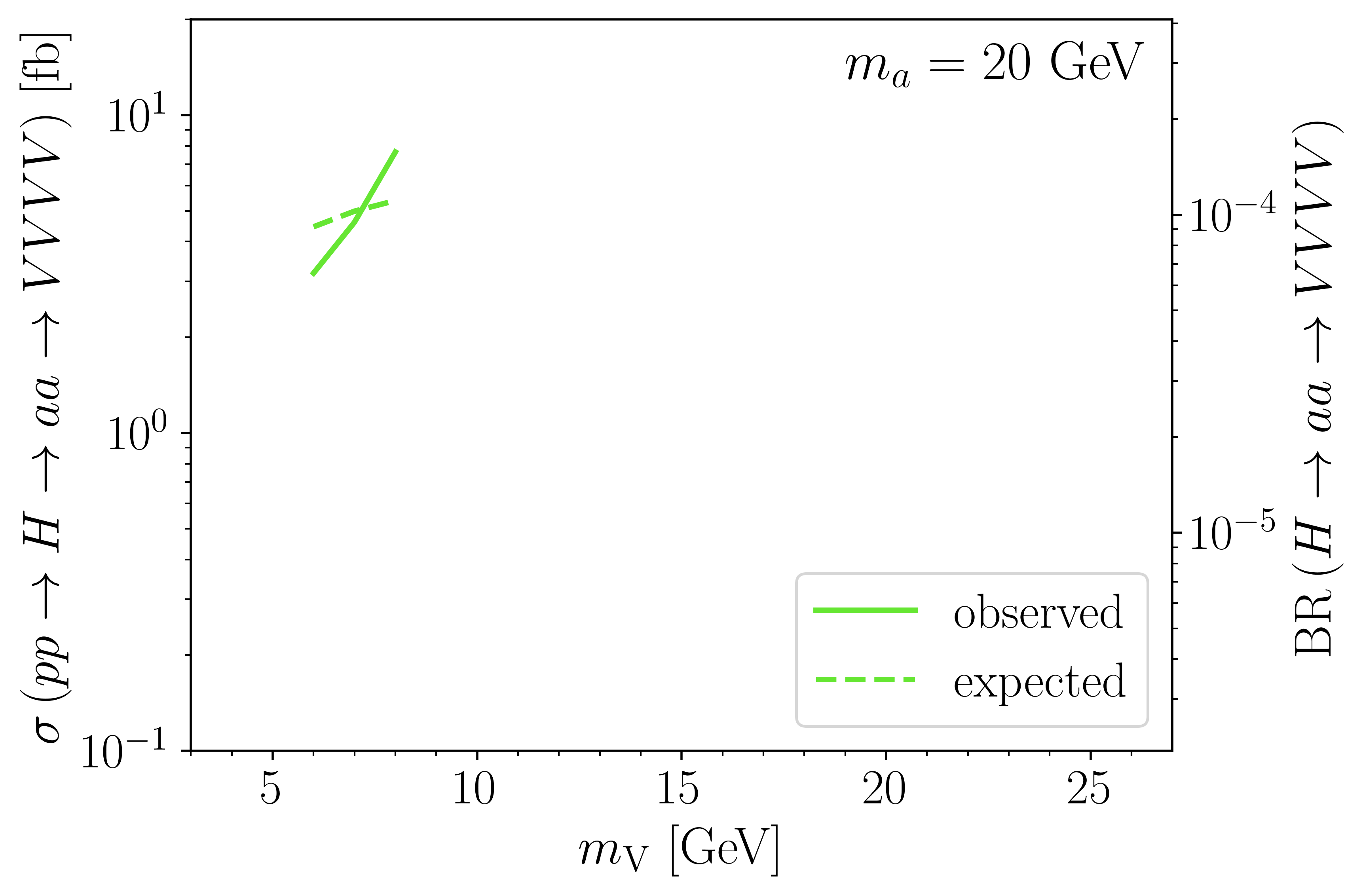}
\includegraphics[width=.45\textwidth]{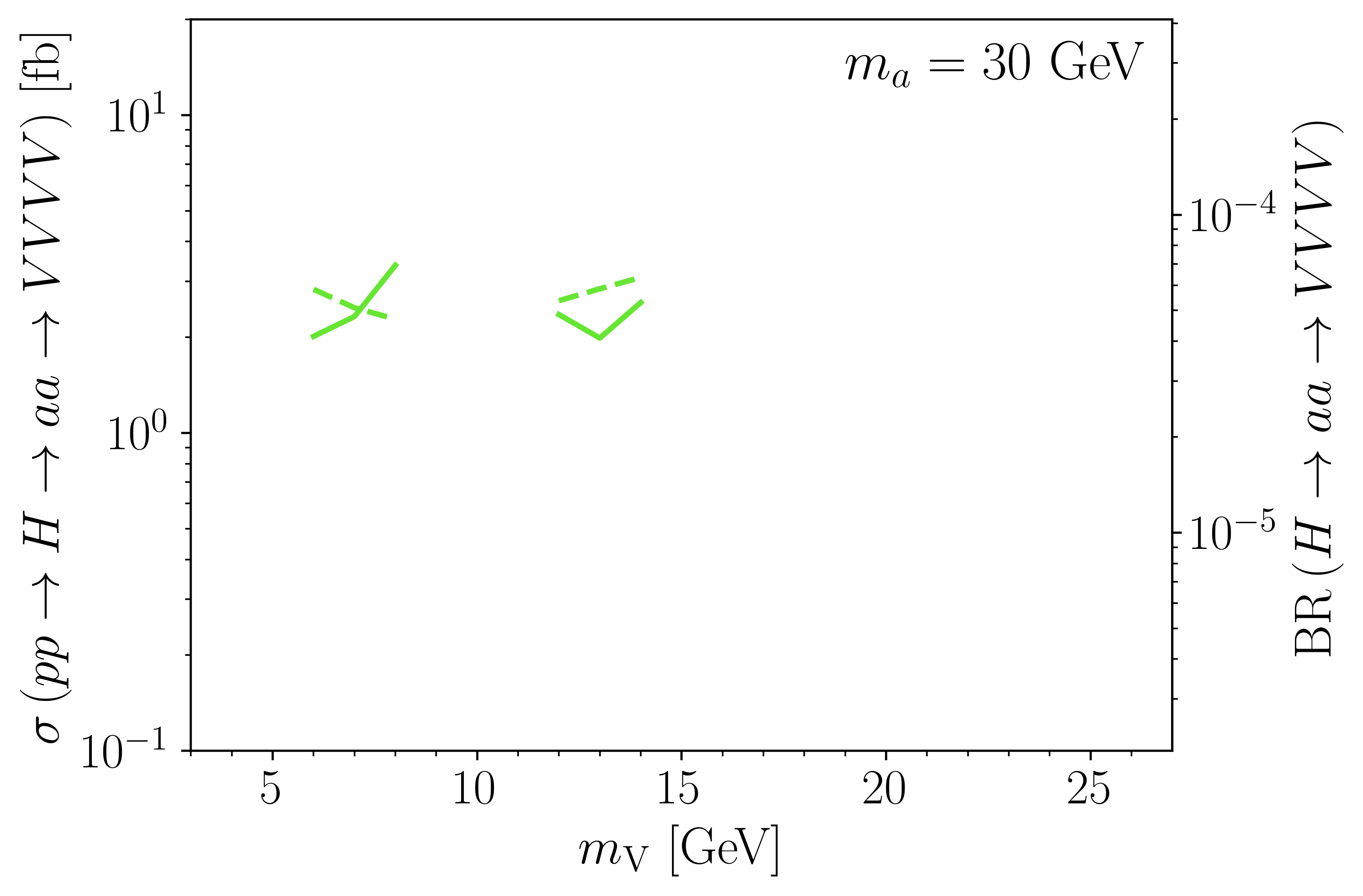}
\includegraphics[width=.45\textwidth]{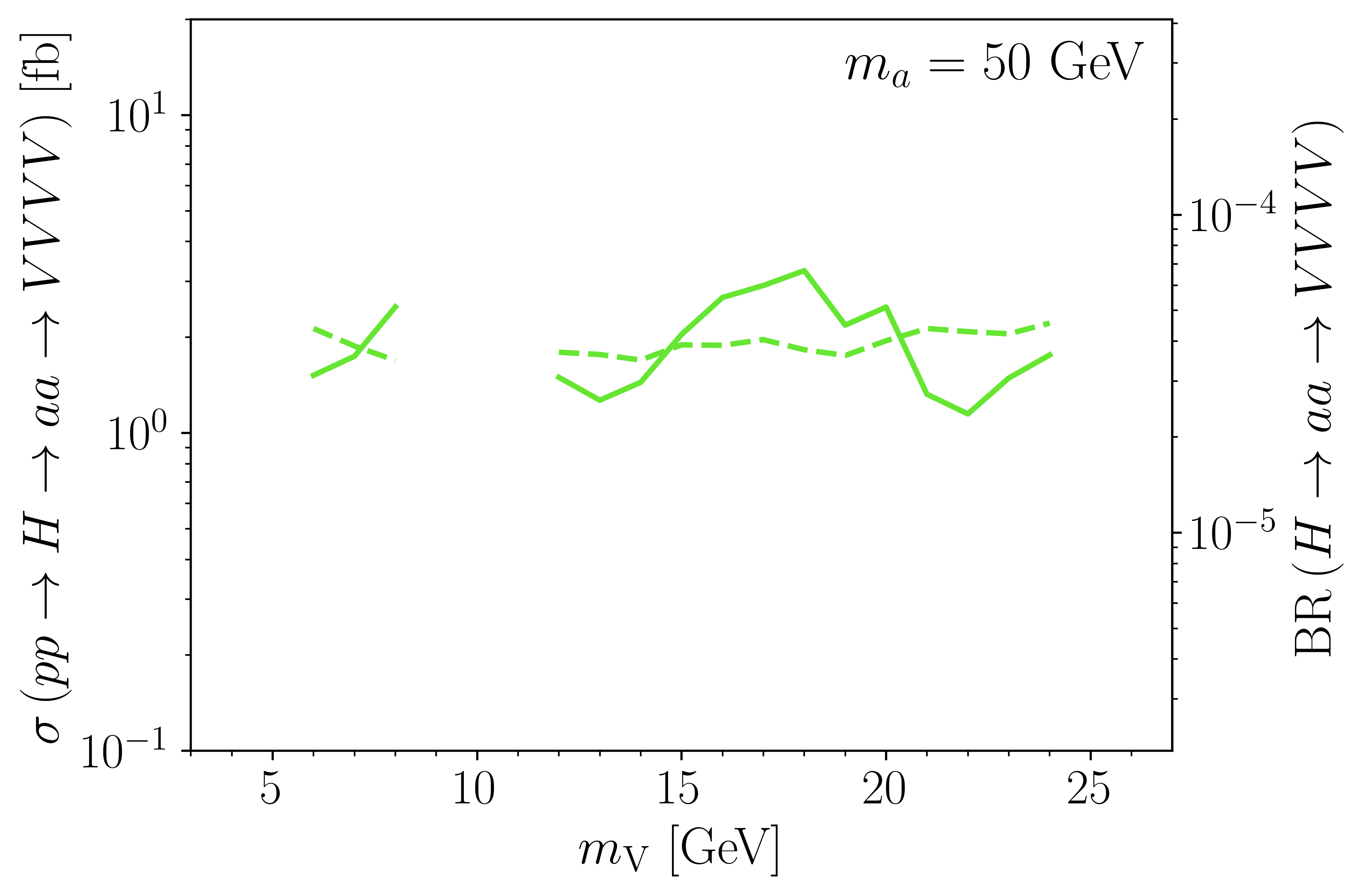}
\caption{Our observed (expected) limits at 95\% confidence level, as a function of $m_V$, on the cross section $\sigma(pp\to H\to aa\to VVVV)$ (left axis) and, assuming the SM value of $\sigma(pp\to H)$, on the Higgs branching fraction BR$(H\to aa\to VVVV)$ (right axis). These limits assume that the $V$ branching fractions are those of a dark photon, as in Fig.~\ref{fig:BR}.}
\label{fig:Higgs8f1}
\end{figure}

Our limits on the $H\to8f$ signature are shown in Fig.~\ref{fig:Higgs8f1}.  Note that we do not include the $V\to\ell\ell$ branching fractions, in contrast to what we and ATLAS have done for the $Z\to6f$ limits. Limits on the branching fraction of $H\to aa\to VVVV$ are strong, of order $4\times 10^{-5}$ -- $1\times 10^{-4}$.   To our knowledge, these are the best limits obtained so far on such a signal.  

 Comparing Figs.~\ref{fig:Z4l1} and \ref{fig:Higgs8f1}, we observe that the latter's limits depend more strongly on $m_a$  than do the former's limits on the corresponding mass $m_{h_D}$. For instance, at $m_V=7$ GeV, limits on $H\to8f$ vary by 2.5 between $m_a=20$ GeV and $m_a=50$ GeV, while for $Z\to 6f$ and $m_{A'}=7$ GeV, limits vary by only 1.4 between $m_a=20$ GeV and $50$ GeV.  The weaker limits at small $m_a$, $m_V$ seem to have multiple sources. First, for the same mass, the $a$  in an $H\to aa$ decay is more boosted than an $h_D$ in a $Z\to h_D A'$ decay, and so its decay products are more collimated.  Leptons from $a$ decays at low $m_a$ are therefore more likely to fail isolation requirements than those from an $h_D$ decay at the same mass.   Second, the same collimation effect makes the decay products of two low-mass $a$ bosons nearly back-to-back. In this case, if the two  $V\to\ell\ell$ pairs in a quadruplet come from two different $a$ decays, their momenta will be in nearly opposite directions.  This can push $m_{4\ell}$ towards its kinematic maximum $m_H-2m_V$, in which case the event may be removed by the  $m_{4\ell}<m_Z-5$ GeV cut.  This effect is less likely at large $m_a$, for which the decay products of the $a$ have a wider angular spread.  

\section{Discussion}

We have reproduced the ATLAS $Z\to 6f$ search of \cite{ATLAS:2023jyp}, and recast it as a limit in the 
 $4\times 10^{-5}$ -- $1\times 10^{-4}$ range on the exotic Higgs decay $H\to aa \to VVVV \to 8f$. One might wonder whether improvements on these limits (other than more data) are possible. There are several directions one might consider.

The worsening limits that we find at lower $m_a$ suggest that in this regime it may be worthwhile to loosen isolation requirements.  Most notably, the ATLAS isolation criteria for each lepton include the $p_T$ or $E_T$ from other leptons. This is important at small $m_a$, where four leptons from a single $m_a$ decay will be relatively collimated. If instead leptons were excluded from each other's isolation variables, the $m_a$ dependence of our results would likely be reduced. It seems unlikely that heavy flavor backgrounds would significantly increase.

Also, loosening the limit on $m_{4\ell}$, while allowing in more background, may also increase efficiency for Higgs and other signals.  The background for $m_{4\ell}>m_Z$ is peaked where the higher-mass dilepton pair has $m_{\ell\ell}\sim m_Z$, whereas the signal has equal and lower-mass dilepton pairs.  

Furthermore, trigger-level analysis at ATLAS (and scouting at CMS) may allow for significantly increased trigger efficiency for signal.  While this obviously lets in more background, most of that background will fail to pass the cuts.

We should note that this is not the first time a recast limit has been attempted on this decay mode. A theory paper \cite{Izaguirre:2018atq} also examined this model and recast a CMS multilepton search where four leptons were required, finding limits on the branching fraction in the $10^{-3}$ range.  
The same paper also proposed that a future search for 5 or more leptons would reach a similar branching fraction limit to the ones we have found here.  However, when we consider their proposal using realistic soft-lepton efficiencies, comparable to those used in in our own analysis, we find weaker limits.  Unfortunately their text is not clear on the efficiencies that they assumed.

Actual limits could be obtained by recasting a recent ATLAS multilepton search \cite{ATLAS:2021yyr} which has a signal region targeting events with $\geq 5$ leptons.  The search is quite inclusive, with trigger paths  similar to the ones used here.  However, the lepton $p_T$ cuts are slightly higher, and the isolation requirements more stringent.  In particular, in addition to the kind of isolation requirements placed in the $Z\to 6f$ analysis, the multilepton analysis removes any lepton within $\Delta R=0.4$ of a jet with $p_T>20$ GeV, and if any two leptons lie within $\Delta R=0.6$ of one another, both leptons are removed unless both have $p_T>30$ GeV.  This last cut substantially reduces sensitivity to the $H\to8f$ signal if $m_V$ is small.  Even at the highest $V$ masses we consider, the expected limits on our signal appear comparable to or weaker than the ones we obtain.  Furthermore, because ATLAS observed an excess in the $\geq5$-lepton channel (21 events over a background of 12.4 $\pm$ 2.3), the observed limits obtained by recasting would be substantially weaker than ours.  Said another way, whereas the multilepton search \cite{ATLAS:2021yyr} sees an excess in its 5 lepton signal region, our recast of the $Z\to6f$ search \cite{ATLAS:2023jyp} disfavors the possibility that it comes from an $H\to8f$ signal with four intermediate $V$ resonances.

We therefore suspect that for signatures with multiple equal-mass dilepton pairs, the $Z\to6f$ search \cite{ATLAS:2023jyp} that we have recast here, which relies on the kinematics of these resonant pairs, tends to be more sensitive than a multilepton-counting search. 
Nevertheless, searches for $\geq 5$ multilepton events have sensitivity to another class of models: those in which leptons are common, but equal-mass dilepton pairs are absent. A recast of this ATLAS search could constrain Higgs decays to off-shell dark photons, cascade decays in HV/DS models with off-shell dilepton pairs, and other scenarios,

The recast we have performed here is particularly straightforward because the $H\to 8f$ signature is simple to simulate and has small theoretical uncertainties.  It is also  important to apply these methods to more complex models, such as confining HV/DS models with dark hadronization.  In such contexts, theoretical uncertainties are potentially much larger, since hadronization in non-QCD-like sectors is poorly understood. We will carry out this recast in a companion paper, where we will argue that large classes of HV/DS models with leptons are excluded by the ATLAS $Z\to 6f$ analysis.

As this paper was in preparation, a new experimental result from CMS appeared \cite{CMS:2024jyb}. Because it requires four muons and imposes higher muon $p_T$ cuts, a first estimate suggests that its limits on the $Z\to 6 f$ and $H \to 8 f$ signals will be substantially weaker than those presented here and in~\cite{ATLAS:2023jyp}.  Also, an ATLAS search appeared that is directly sensitive to $h_d\to A'A'$ and $a\to VV$ decays \cite{ATLAS:2024zoq}. But again its $p_T$ cuts are higher than those in \cite{ATLAS:2023jyp}, reducing its sensitivity to the $Z\to 6 f$ and $H \to 8 f$ signals. 

\

\

\section*{Acknowledgements}
We thank Melissa Franklin, Jerry Ling, Mingyi Liu, Aaron White and Lailin Xu for useful conversations. JC and RH are supported in part by the DOE Grant DE-SC-0013607. RH was also supported by the NSF GRFP fellowship. LL is supported by the DOE Grant DE-SC-0010010.  MJS thanks Harvard University for its hospitality during this research.

\bibliographystyle{JHEP}
\bibliography{Reference}
\end{document}